\documentclass[reprint,letter,floatfix,aps,twocolumn, noeprint,superscriptaddress,notitlepage,linenumbers]{revtex4-2}

\usepackage{graphicx} 
\usepackage{epsfig}
\usepackage{epstopdf}
\usepackage{amsmath}
\usepackage{amsfonts}
\usepackage{amssymb} 
\usepackage{color}
\usepackage{multirow}
\usepackage{lineno}
\usepackage{url}
\usepackage{soul}
\usepackage[normalem]{ulem}
\usepackage[dvipsnames]{xcolor}
\usepackage{balance}
\usepackage{textcomp} 
\graphicspath{{figures/}}

\begin{document}
\nolinenumbers

\title{Deep exclusive electroproduction of $\pi^0$ at high $Q^2$ in the quark valence regime} 
\author{	M.~Dlamini	} \affiliation{	Ohio University, Athens, Ohio 45701, USA	}
\author{	B.~Karki	} \affiliation{	Ohio University, Athens, Ohio 45701, USA	}	
\author{	S.F.~Ali	} \affiliation{	Catholic University of America, Washington, DC 20064, USA}	
\author{	P-J.~Lin	} \affiliation{Universit\'e Paris-Saclay, CNRS/IN2P3, IJCLab, 91405 Orsay, France}
\author{	F.~Georges	} \affiliation{Universit\'e Paris-Saclay, CNRS/IN2P3, IJCLab, 91405 Orsay, France}
\author{	H-S~Ko	} \affiliation{Universit\'e Paris-Saclay, CNRS/IN2P3, IJCLab, 91405 Orsay, France}\affiliation{Seoul National University, 1 Gwanak-ro, Gwanak-gu, 08826 Seoul, Korea}
\author{	N.~Israel	} \affiliation{	Ohio University, Athens, Ohio 45701, USA	}
\author{	M.N.H.~Rashad	} \affiliation{	Old Dominion University, Norfolk, Virginia 23529, USA	}	
\author{	A.~Stefanko	} \affiliation{	Carnegie Mellon University, Pittsburgh, Pennsylvania 15213, USA	}
\author{	D.~Adikaram} \affiliation{	Thomas Jefferson National Accelerator Facility, Newport News, Virginia 23606, USA	}			
% ...
\collaboration{The Jefferson Lab Hall A Collaboration}

\date{\today}

\begin{abstract}

We report measurements of the exclusive neutral pion electroproduction cross section off protons at large values of $x_B$ (0.36, 0.48 and 0.60) and $Q^2$ (3.1 to 8.4 GeV$^2$) obtained from Jefferson Lab Hall A experiment E12-06-014.  The corresponding structure functions $d\sigma_T/dt+\epsilon d\sigma_L/dt$, $d\sigma_{TT}/dt$, $d\sigma_{LT}/dt$ and $d\sigma_{LT'}/dt$ are extracted as a function of the proton momentum transfer $t-t_{min}$. 
The results suggest the amplitude for transversely polarized virtual photons continues to dominate the cross section throughout this kinematic range. The data are well described by calculations based on  transversity Generalized Parton Distributions coupled to a helicity flip Distribution Amplitude of the pion, thus providing a unique way to probe the structure of the nucleon. 
\end{abstract}

\maketitle

\begin{figure}[b]
\centering\includegraphics[width=0.9\linewidth]{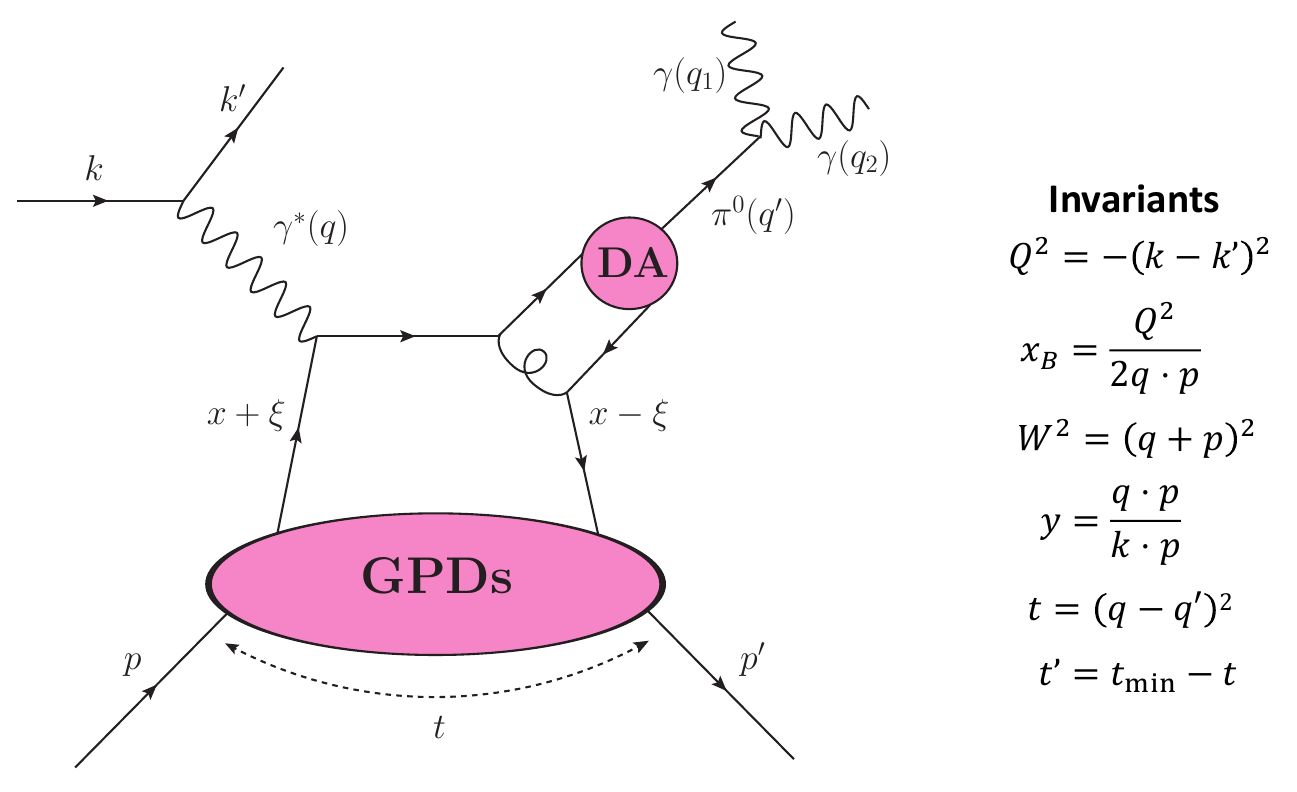}
\caption{Leading twist diagram representing the pseudoscalar DVMP to the $\gamma\gamma$ channel. 
The net four-momentum transferred to the proton is $t$, whose minimum value $t_{min}$ occurs when the $\pi^0$ meson is emitted parallel to the virtual photon. The average light-cone momentum fraction carried by the struck parton is $x$ with $-2\xi$  the light-cone momentum transfer. 
}
\label{HandbagFig}
\end{figure}

Generalized Parton Distributions (GPDs)~\cite{Ji:1996ek,Mueller:1998fv,Radyushkin:1997ki} describe the three-dimensional structure of the nucleon by correlating the transverse position and the longitudinal momentum of the quarks and gluons inside of it. GPDs are accessible through deep exclusive processes, such as Deeply Virtual Compton Scattering (DVCS) and Deeply Virtual Meson Production (DVMP).
For the latter, collinear factorization theorems~\cite{Collins:1996fb} applied to longitudinally polarized virtual photons only (not to the transversely polarized ones) establish that the DVMP amplitude  factorizes at large $Q^2$ into a hard perturbative part and a soft component described by the GPDs of the nucleon.
Figure~\ref{HandbagFig} shows the leading mechanism of the $\pi^0$ electroproduction reaction and defines the kinematic variables of the process. There are four chiral-even GPDs 
($H,\, E,\, \widetilde H, \widetilde E$) that define the quark helicity-conserving amplitudes and four chiral-odd (transversity) GPDs ($H_T,\, E_T,\, \widetilde H_T, \widetilde E_T$) that define the quark helicity-flip amplitudes. In the Bjorken limit where $Q^2 \rightarrow \infty$, the target rest-frame energy of the virtual photon  $\nu \rightarrow \infty$ and ${t}/{Q^2} \ll 1$,  QCD  predicts that the reaction cross section is dominated by the contribution of longitudinally polarized virtual photons. This longitudinal component depends on the momentum transfer as $Q^{-6}$, whereas the transverse component goes asymptotically as $Q^{-8}$. The longitudinal cross section of DVMP only depends on the convolution of  chiral-even GPDs of the nucleon with the quark helicity-conserving Distribution Amplitude (DA)  of the meson~\cite{Lepage:1980fj}. However, existing data~\cite{Masi_2008, Fuchey:2010kna, Bedlinskiy:2012be,PhysRevC.90.025205, Defurne:2016eiy, Mazouz:2017skh,KIM2017168, Alexeev:2019qvd} for neutral pseudoscalar meson production in the quark valence regime, with limited reach in $Q^2$,  show that transversely polarized virtual photons   dominate the total cross section. 
In the collinear approximation, singularities occur for transversely polarized photons and mesons. 
To explain the large transverse contribution to the  $\pi^0$ electroproduction cross sections, it has been suggested~\cite{PhysRevD.79.054014,Goloskokov:2011rd, Goldstein_2012} to regularize these singularities by including transverse degrees of freedom of the quarks and anti-quarks making up the meson. In this framework, the $\pi^0$ electroproduction cross section is described by the convolution of a higher order helicity-flip DA of the meson with the transversity GPDs of the nucleon. Calculations based on this approach~\cite{PhysRevD.79.054014,Goloskokov:2011rd} were able to reproduce reasonably well the existing neutral pseudoscalar meson production data cited above.  This letter reports measurements of $\pi^0$ electroproduction cross sections that extend to higher 
values of $Q^2$ (from 3.1 to 8.4 GeV$^2$) and of $x_B$ (0.36, 0.48 and 0.60), with a large coverage in $t$ and center of mass energy $s$. 

\begin{table*}
  \begin{tabular}{cccccccccc}
    \hline \hline
  
    $x_B$-label & \multicolumn{3}{c}{ 0.36} & \multicolumn{4}{c}{0.48} & \multicolumn{2}{c}{0.60} \\
    $\langle x_{B}\rangle$ & $\qquad$ 0.36 & 0.36 & 0.36 $\qquad$ & $\qquad$ 0.48 & 0.45 & 0.46 & 0.46 $\qquad$&$\qquad$ 0.59 & 0.60 $\qquad$\\
    $E$  (GeV) &$\qquad$ 7.38 & 8.52 & 10.59 $\qquad$&$\qquad$ 4.49 & 8.85 & 8.85 & 10.99 $\qquad$&$\qquad$ 8.52 & 10.59$\qquad$\\
    $Q^2$ (GeV$^2$) &$\qquad$3.11 & 3.57 & 4.44 $\qquad$&$\qquad$ 2.67 & 4.06 & 5.16 & 6.56 $\qquad$&$\qquad$ 5.49 & 8.31$\qquad$\\
    $W^2$ (GeV$^2$) &$\qquad$ 6.51 & 7.29 & 8.79 $\qquad$&$\qquad$ 3.81 & 5.62 & 6.67 & 8.32 $\qquad$&$\qquad$ 4.58 & 6.46$\qquad$\\
    $-t_{min}$ (GeV$^2$) &$\qquad$ 0.16 & 0.17 & 0.17 $\qquad$&$\qquad$ 0.33 & 0.35 & 0.35 & 0.36 $\qquad$&$\qquad$ 0.67 & 0.71$\qquad$\\
    $\epsilon$&$\qquad$ 0.61 & 0.62 & 0.63 $\qquad$&$\qquad$ 0.51 & 0.71 & 0.55 & 0.52 $\qquad$&$\qquad$ 0.66 & 0.50$\qquad$\\
    \hline \hline
  \end{tabular}
  \caption{Incident beam energy $E$ and average values for scattering kinematic variables for each of the nine ($E, Q^2, x_B$) settings where the $\pi^0$ cross sections are reported. For each setting, cross sections are measured as a function of $t'=t_{min}-t$,
  with $t_{min}$ calculated event-by-event. 
  }
  \label{tab:kin}
\end{table*}

The exclusive meson electroproduction cross section can be written ~\cite{Drechsel:1992pn}  in terms of contributions from longitudinally (L) and transversely (T) polarized photons and their interference as:
\begin{multline}\frac{d^4\sigma}{dQ^2 dx_B dt d\phi} =   \frac{1}{2\pi}\frac{d^2~ \Gamma_{\gamma}}{dQ^2 dx_B}(Q^2,x_B,E)\\
\Bigg[ \frac{d\sigma_{\mathrm{T}}}{dt}  + \epsilon \frac{d\sigma_{\mathrm{L}}}{dt} +
   \sqrt{2\epsilon (1+\epsilon )}\frac{d\sigma_{\mathrm{LT}}}{dt}\cos(\phi)+
     \epsilon \frac{d\sigma_{\mathrm{TT}}}{dt}\cos{(2\phi)} \\
     +h\sqrt{2\epsilon (1-\epsilon)}\frac{d\sigma_{\mathrm{LT'}}}{dt}
  \sin(\phi) \Bigg],
 \label{DVMPCrossEquation}
\end{multline}
where $h(\pm 1)$ is the helicity of the initial lepton, $E$ is the incident beam energy and $\phi$ is an angle between leptonic and hadronic planes defined according to the Trento convention~\cite{Bacchetta:2004jz}. The virtual photon flux  \cite{PhysRev.129.1834} $\frac{d^2~ \Gamma_{\gamma}}{dQ^2 dx_B} $ and the degree of longitudinal polarization  $\epsilon$ are  defined as: 
    \begin{equation}
     \frac{d^2~ \Gamma_{\gamma}}{dQ^2 dx_B}(Q^2,x_B,E)=\frac{\alpha}{8\pi}~\frac{1}{1-\epsilon} ~\frac{1-x_B}{x_B^3} ~ \frac{Q^2}{M_p^2E^2}~ 
   \end{equation}
   \begin{equation}
\epsilon = \frac{1-y-\frac{Q^2}{4E^2}}{1-y+\frac{y^2}{2}+\frac{Q^2}{4E^2}}
   \end{equation}  
where $M_p$ is the proton mass and  $y=[q\cdot p]/[k\cdot p]$.

Experiment E12-06-114 took data between 2014 and 2016 in Jefferson Lab Hall A. The main goal of this experiment was to measure the DVCS cross section $ep\rightarrow ep\gamma$. The same experimental configuration also captured exclusive $\pi^0$ electroproduction events. The kinematics covered by the experiment are shown in Tab.~\ref{tab:kin}. The electron beam scattered off a 15-cm-long liquid hydrogen target with luminosities greater than $10^{38}$~cm$^{-2}$s$^{-1}$. The beam polarization measured by the Hall A M\o ller polarimeter was $86\pm 1\%$, with the uncertainty dominated by the systematic precision of the measurement. Scattered electrons were detected in a High-Resolution Spectrometer (HRS) with a relative momentum resolution of $2\cdot 10^{-4}$ and a horizontal angular resolution of 2 mr~\cite{Alcorn:2004sb}. Photons from the DVCS and DVMP processes were measured in an electromagnetic calorimeter consisting of a 13$\times$16 array of  $\mathrm{PbF_2}$ crystals. The analog signal of each channel was sampled by a 1~GHz Analog Ring Sampler~\cite{Feinstein:2003vi,Druillole:2001dm} and recorded over 128~ns. The calorimeter was calibrated multiple times during the experiment using coincident elastic H(e, $e'_\text{Calo}$ $p_\text{HRS}$) events. The typical energy resolution of the calorimeter was 3\% at 4.2~GeV with an angular resolution of 1.5 mr (when located 6~m  from the target). Between two consecutive elastic calibrations, the output of the calorimeter for a given photon energy changed up to 10\% due to the radiation damage of the $\mathrm{PbF_2}$ crystals. The loss of signal was estimated and compensated for by adjusting the reconstructed invariant mass of the detected $\pi^0$ events.  Additional details are presented in \cite{SMref1}.

Neutral pions were reconstructed by selecting 2 photons in the calorimeter above 500~MeV each, in coincidence with the detection of a scattered electron in the HRS. 
The HRS-calorimeter coincidence-time resolution was about 1~ns. 
The total contribution from accidental coincidences was below 2\% and was subtracted from the experimental yield.  The $\pi^0$ sample was cleanly identified by selecting events around the  invariant mass $m_{\gamma\gamma}=\sqrt{(q_1+q_2)^2}$. 
 The exclusivity of the reaction was ensured by reconstructing the missing-mass squared $M_X^2$ of the $H(e,e'\gamma \gamma)X$ reaction (see figure in the supplemental material \cite{SMref2}). 
 
\begin{figure*}[t]
\centering

\vspace{-2.0cm}

\includegraphics[width=0.75\textwidth,angle=-90]{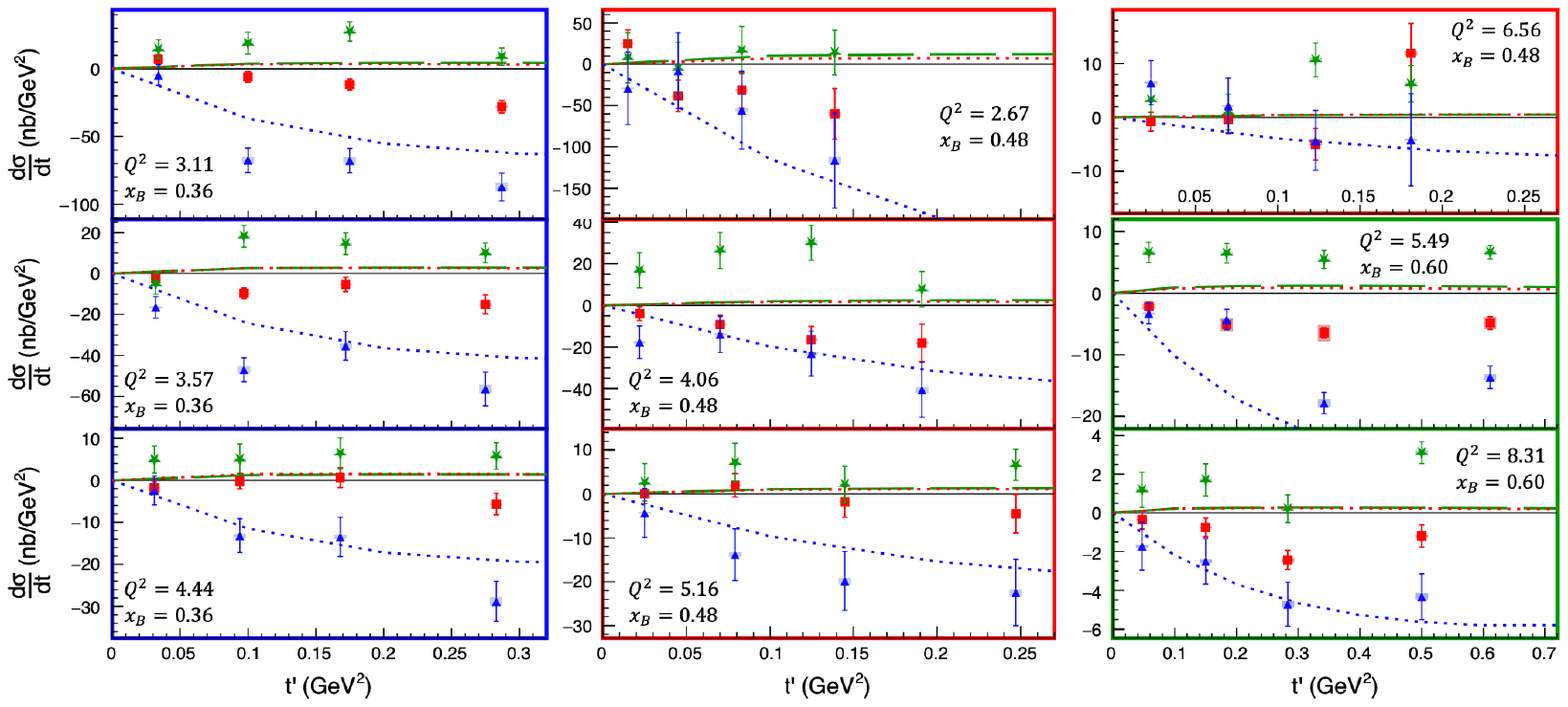}

\vspace{-3cm}

\caption{Structure functions $d\sigma_{TT}$(blue triangles), $d\sigma_{LT}$(red squares) and $d\sigma_{LT'}$ (green stars) for all kinematic setting as a function of $t'=t_{min}-t$. The dashed curves are calculations based on transversity GPDs of the nucleon~\cite{Goloskokov:2011rd}. The gray boxes surrounding the data points show the systematic uncertainty. }
  \label{cross-sectionHalfFig1}
\end{figure*}

The acceptance and resolution of the experiment were computed by a Monte-Carlo simulation based on the {\sc Geant4} software~\cite{Agostinelli:2003}. The simulation and cross
section extraction includes the real and virtual radiative effects,
based on calculations 
of~\cite{Vanderhaeghen:2000ws}, see also Supplemental Material \cite{SMref3}.
 
Data were binned into 12 $\phi$  bins by 5 $t'$ bins. The different structure functions appearing in the  $\pi^0$ electroproduction cross section were extracted by exploiting their specific $\phi$ dependencies, minimizing the $\chi^2$ between the number of experimental and simulated events: 
\begin{equation}
\label{Bchi}
\chi^{2}= \sum_{i=1}^{\mathcal{N}=60}\left( \frac{N_i^{exp} - N_i^{sim}}{\sigma_i^{exp}}\right)^{2}
\end{equation}
where the sum runs over all $12\times5$ bins for each ($x_B, Q^2$) setting. $N_i^{exp}$ is the total number of events in bin $i$ with corresponding statistical precision $\sigma_i^{exp}$. The number of simulated events in bin $i$ is computed by convoluting the acceptance and resolution of the experimental setup with the kinematic dependencies of each of the  structure functions ($d\sigma_T/dt+\epsilon d\sigma_L/dt$, $d\sigma_{TT}/dt$ and $d\sigma_{LT}/dt$) that make up the cross section (see Eq. \ref{DVMPCrossEquation}). These structure functions are the free parameters of the $\chi^2$ minimization. An example of these fits and the numerical values all the extracted structure functions  are shown in the supplemental material \cite{SMref4}. The helicity-dependent structure function $d\sigma_{LT'}$ is extracted by a similar fit to the difference in yield for events with opposite helicities. Bin migration effects from one kinematic bin to another due to resolution and radiative effects are incorporated into the simulation and are  up to 10\% depending upon the kinematic bin. Cross sections are only reported for the 4 lowest $t'$ bins; the additional highest $t'$ bin in the analysis is only used to evaluate bin migration to the lower $t'$ bins.  The
systematic uncertainty associated with the bin migration is assessed by varying the selection cut
on the missing mass-squared, for each kinematic bin. The $d\sigma_T/dt+\epsilon d\sigma_L/dt$, $d\sigma_{TT}/dt$ and $d\sigma_{LT}/dt$ values extracted from the fit show a degree of correlation of around 10\% at low $t'$, but this correlation reaches 90\% at large $t'$ due to the loss of full azimuthal acceptance in the detector.

The total systematic uncertainty of the results reported herein varies between 4\% and 8\% depending on the kinematic setting. The variation in the systematic uncertainty from one setting to another is due to the effect of the exclusivity cut, which is very sensitive to our ability to reproduce in the simulation the actual energy resolution of the photons as a function of their impact position onto the calorimeter.

Figure~\ref{cross-sectionHalfFig1} shows the measurements of the structure functions $d\sigma_{TT}$, $d\sigma_{LT}$ and $d\sigma_{LT'}$ at the kinematics settings listed in Tab.~\ref{tab:kin}. In general $d\sigma_{TT}$ is larger that the interference terms involving the longitudinal amplitude ($d\sigma_{LT}$ and $d\sigma_{LT'}$). This hints at a dominance of the transverse amplitude in the reaction mechanism. Data are compared to calculations from the modified factorization approach first introduced in~\cite{PhysRevD.79.054014,Goloskokov:2011rd}. This model provides a large contribution to the transverse amplitude which arises from the convolution of  chiral-odd (transversity) GPDs of the nucleon with a quark-helicity flip pion DA, whereas the longitudinal amplitude is extremely small, as illustrated by the calculations of $d\sigma_{LT}$ and $d\sigma_{LT'}$ in this framework. It is interesting to note that the data show a stronger longitudinal amplitude than in the model, which underestimate the values of both $d\sigma_{LT}$ and $d\sigma_{LT'}$, while providing a good agreement with $d\sigma_{TT}$. The underestimation of d$\sigma_{LT'}$ was already observed in~\cite{Masi_2008, KIM2017168}. The sign of the interference structure function $d\sigma_{LT}$ is measured to be systematically opposite to the theory calculations. 
In these model calculations of $d\sigma_{LT}$, the contributions
from the real parts of $H_T$ and $\widetilde E$ on one hand, and 
${\overline E}_T = 2{\widetilde H}_T+E_T$ and $\widetilde H$ on the other hand enter with 
opposite sign.  The latter term is small, and therefore these data for 
$d\sigma_{LT}$ will strongly constrain models of the currently poorly known GPD $\widetilde E$.

\begin{figure}[t!]
\includegraphics[width=\linewidth, bb= 25 55 372 615, clip]{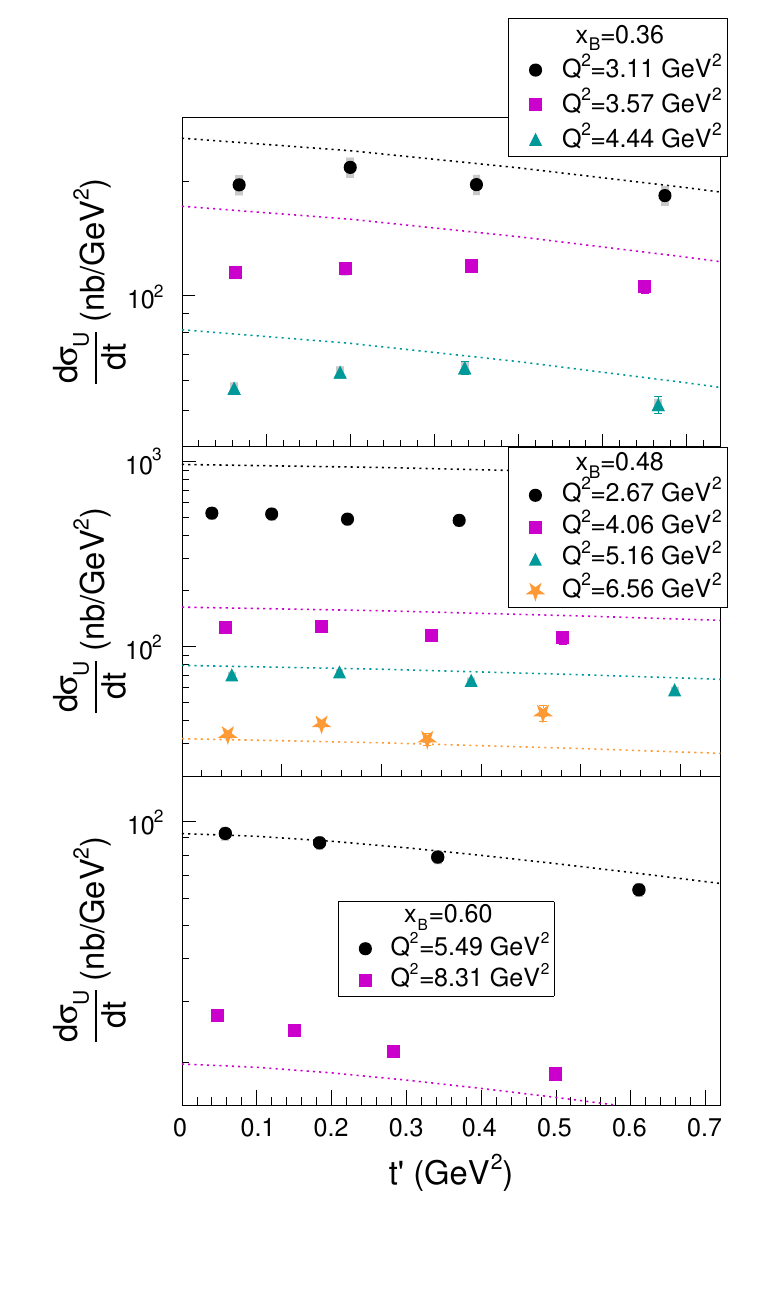}
\caption{Structure function $d\sigma_U=d\sigma_{T} + \epsilon d\sigma_{L}$ as a function of $t'=t_{min}-t$ for all kinematic settings.  The gray boxes surrounding the data points show the systematic uncertainty. The dashed curves are calculations which include (and are dominated by) transversity GPDs of the nucleon}~\cite{Goloskokov:2011rd}. 
\label{cross-sectionHalfFig2}
\end{figure}

Figure~\ref{cross-sectionHalfFig2}  shows the measurements of the unpolarized  structure function $d\sigma_U=d\sigma_T+\epsilon d\sigma_L$. Calculations based on the modified factorization approach~\cite{Goloskokov:2011rd} are in reasonable agreement. This has been observed at lower values of $Q^2$ ($<3$~ GeV$^2$)~\cite{Bedlinskiy:2012be,PhysRevC.90.025205,Defurne:2016eiy}. The fact that this is still true at these much higher values of the momentum transfer indicates that the asymptotic regime predicted by QCD, where the longitudinal amplitude must dominate, is not yet reached. On the other hand, %
at the highest value of $Q^2$ the transverse dominated calculations underestimate the data, thus providing some evidence of a sizeable longitudinal contribution, as also confirmed in Fig.~\ref{cross-sectionHalfFig1} by the fact that $d\sigma_{LT}$ is becoming relatively larger compared to $d\sigma_{TT}$. 

\begin{figure}[!b]

\centering\includegraphics[width=1\linewidth, bb=5 37 532 612, clip]{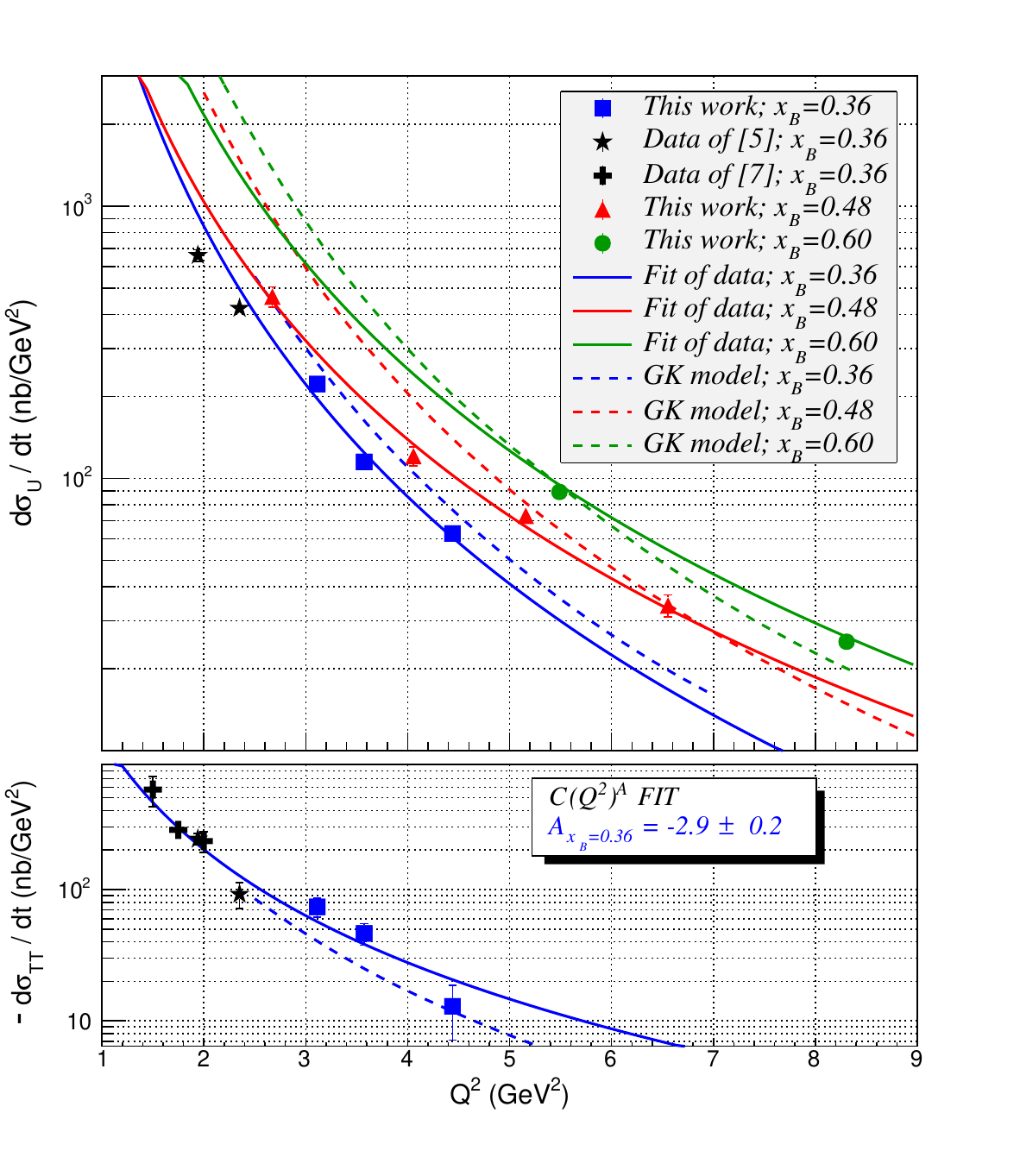}
\caption{The $Q^2$ dependence of the structure functions $d\sigma_U$ and $-d\sigma_{TT}$ at $\langle{}t' \rangle{}=0.1$~GeV$^2$. The closed markers are the experimental results, the solid curves are the fitted functions, and the dotted curves are the  predictions of  \cite{Goloskokov:2011rd}. The bars on the closed markers show their statistical and systematic uncertainties added in quadrature. The $d\sigma_U$ and $-d\sigma_{TT}$ from this experiment and the corresponding curves at the settings $x_B=0.36$, $0.48$, and $0.60$ are shown in blue, red, and green respectively. The black stars and crosses show the results from \cite{Fuchey:2010kna} and \cite{Defurne:2016eiy} correspondingly,  which are also included in the fit at $x_B=0.36$. }
  \label{cross-sectionFig}
\end{figure}

The $Q^2-$dependence of the structure functions is particularly interesting to study, as its asymptotic limit is the only feature that can be predicted from first principles (i.e. QCD) for different reaction mechanisms. 
Figure~\ref{cross-sectionFig} (top) shows the $Q^2-$dependence of $d\sigma_U=d\sigma_T+\epsilon d\sigma_L$  at constant $t'=0.1$~GeV$^2$ and all three  values of $x_B$. 
A broader perspective on the $Q^2$- and $t$-dependence of these results is presented by the fits in Tab.~\ref{tbl:GlobalFit}.  At each $x_B$ setting, we fit the data to a functional form
$C (Q^2)^A \exp(-Bt')$.  These fits, plotted in Fig.~\ref{cross-sectionFig} at
fixed $t'$
demonstrate an approximately global $1/Q^6$ behavior of the cross section over the  $t'$ and $x_B$ range.
The calculations based on the modified factorization approach show a 
steeper variation with $Q^2$ (approximately $Q^{-7}$) than 
the dependence observed in the data. This suggests a more significant longitudinal component in the data than in the model, which is also compatible with the significantly larger values $d\sigma_{LT}$ shown in Fig.~\ref{cross-sectionHalfFig1}. The bottom panel in Fig.~\ref{cross-sectionFig} shows the $Q^2-$dependence of $d\sigma_{TT}$ which is also 
incompatible with the asymptotic limit $\sim Q^{-8}$.

Figure~\ref{cross-sectionFig} also shows the comparison with the previous available data at lower $Q^2$ and illustrates the much higher reach of these new measurements to best constrain the $Q^2-$dependence of the cross section, and for different values of $x_B$. These data also reach large values of $t=t_{min}-t'$, with the central value of $-t$ up to 1.3 GeV$^2$. The $t-$dependence of the cross section, often parametrized by Regge-like profile functions, is no longer valid at typical values of $-t>1$~GeV$^2$. This was realized in the GPD analysis of nucleon form factors~\cite{Diehl:2013xca}. The theory calculations shown herein include a profile function with a strong $x\otimes t$ correlation~\cite{Kroll:2017hym}, which also allows the proton radius to remain finite as $x\to 1$ and allows the proton form factors---the lowest moments of GPDs--to behave as powers of $t$ at large $-t$. One must point out, though, that these calculations are obtained using some kinematic approximations, such as $\xi\approx x_B/(2-x_B)$. Recent theory  developments~\cite{Braun:2014} have shown that power corrections of $\mathcal O(t/Q^2)$ and $\mathcal O(M_p/Q^2)$ should be included and recent DVCS data~\cite{Defurne:2017paw} at similar kinematics have been proved sensitive to these effects.


The longitudinal to transverse ratio $R$ of exclusive $\rho^0$ electro-production was measured at HERA over the range of $Q^2$ from $\le 1\text{ GeV}^2$ to $\ge 20 \text{ GeV}^2$ \cite{Chekanov:2007zr,Aaron:2009wg}.  Over this kinematic range, $R$ rises
from $\approx 1$ to $\approx 5$ as $Q^2$ increases. Thus even at $Q^2 \sim 20 \text{  GeV}^2$ the transverse cross section in deep virtual exclusive $\rho$ production is not negligible. The role of the
pion as the Goldstone boson of Chiral Symmetry breaking predicts a much
smaller value of $R$ for exclusive $\pi^0$ production for $Q^2$ in the
range of 1 to 3 GeV$^2$ \cite{PhysRevD.79.054014,Goloskokov:2011rd, Goldstein_2012}.  Nonetheless we expect a gradual transition to dominance of $\sigma_L$ in $\pi^0$ electroduction as $Q^2$ increases. Observing this transition is crucial
to disentangling the contributions of quark helicity-flip and helicity-conserving
amplitudes.
The present
data demonstrate slower than asymptotic $Q^2$-dependence and also
provide initial evidence
for the interference of quark helicity-flip and helicity-conserving
amplitudes in $d\sigma_{LT}$. 
An L/T separation of the $\pi^0$ electroproduction cross section at these high values of $x_B$ will provide a definite answer on the size of the longitudinal contribution. This is the goal of an upcoming experiment~\cite{Munoz:2013} in Hall C at Jefferson Lab which is expected to run within the next two years.\\

\begin{table}[t]
\centering
\caption{Combined $(Q^2,t')$ fits $d\sigma_U =C (Q^2)^A \exp(-Bt')$at each $x_B$ setting. Only the data of this publication
are included. The fits and error bars are based on the statistical and systematic uncertainties of the data, added in quadrature.}
\begin{tabular}{cccccr}
\hline\hline
$x_B$ & $C$ & $A$ & $B$ & $\chi^2$ & \textnumero \\
 & $\mu$b/GeV$^2$ &  & $\text{GeV}^{-2}$ & Total &\ d.o.f.\\
\hline
\ 0.36 \ & $8.6\pm 1.4$ & $-3.3 \pm 0.1$ & $0.34\pm 0.17$ & 18. & 9\\
0.48 & $8.3\pm 0.9$& $\ -2.9\pm 0.1\ $ & $0.69 \pm 0.3$ & 27. & 13\\
0.60 &$20. \pm 4.$ & $-3.1\pm 0.1$& $0.75 \pm 0.1$ & 1.6 & 5\\
\hline
\end{tabular}
\label{tbl:GlobalFit}
\end{table}

\section*{Acknowledgements}
We thank P. Kroll and S. Liuti, 
 for very fruitful discussions about these results. We acknowledge essential contributions by the Hall A collaboration and Accelerator and Physics Division staff at Jefferson Lab. This material is based upon work supported by the U.S. Department of Energy, Office of Science, Office of Nuclear Physics under contract DE-AC05-06OR23177. This work was also supported a DOE
Early Career Award to S. Covrig Dusa for  the development of
high power hydrogen target cells,
the National Science Foundation (NSF), the French CNRS/IN2P3, ANR,  and P2IO Laboratory of Excellence, and  the Natural Sciences and Engineering Research  Council of Canada (NSERC).\\

\vspace{15cm}


\bibliography{biblography}

\onecolumngrid
\clearpage
\twocolumngrid

\clearpage
\section{Supplemental Material}
\vskip -1em

\subsection{Radiative Corrections}
The radiative corrections are calculated in three parts.
The emission of real photons is included in the simulation, in the peaking approximation.  The integration of the simulation over the experimental acceptance automatically incorporates the correction for  events radiated out of the exclusivity window.
The finite part of the virtual radiative corrections are
calculated with the code of Vanderhaeghen \textit{et al} \cite{Vanderhaeghen:2000ws}.
The code includes the coherent sum of the BH and VCS
amplitudes.  For the calculations in this work, the BH amplitude
is set to zero, therefore the code produces a general
electroproduction  radiative correction.  In  \cite{Vanderhaeghen:2000ws} Eq.~58, there is an approximately constant term (weakly dependent on cut-off) from real radiation 
\begin{align}
\delta_R^{(0)}  
   &=
   \frac{\alpha}{\pi}\left[\frac{1}{2} \ln^2\frac{Q^2}{m_e^2}-\ln^2\frac{\widetilde E'_e}{\widetilde E_e} 
     - \frac{\pi^2}{3} + \text{\tt Sp}\left(\widetilde u^2\right) \right]
\end{align}
with {\tt Sp} the Spence function of $u^2=\cos^2(\theta_e/2)$ and $\theta_e, E'_e$ the electron scattering angle and energy.  These tilded-variables are evaluated
in the CM system of the recoiling proton plus radiated (second) photon.
We compute this term at the average $\Delta^2=t$ for
each $(Q^2,x_B)$ kinematic at the cutoff value
of $M_X^2$. These correction factors are listed in 
Table~\ref{tbl:RadCorr}.  We do not find any variation of these
correction factors within the $t$-ranges of each setting.

\begin{table}[bp]
\centering
\caption{Radiative Correction Factors}
\begin{tabular}{ccccccc}
\hline\hline
Kin & $x_B$ & $Q^2 (\text{GeV}^2)$ & $E$  (GeV) & $\exp(-\delta_R(0))$ & Virtual & Total \\
\hline
$36\_1$ & 0.36 & 3.20 & 7.38 & $0.737$ & $1.276$ & 0.941 \\ 
36\_2 &        & 3.60 & 8.517 & 0.734   &   1.283 & 0.941 \\
36\_3 &     &    4.47 & 10.617& 0.727   &   1.294 & 0.941 \\
\hline
48\_1 & 0.48 & 2.70  & 4.483 & 0.742 & 1.269 & 0.942 \\
48\_2 &      & 4.365 & 8.843 & 0.728 & 1.292 & 0.941 \\
48\_3 &      & 5.334 & 8.843 & 0.723 & 1.302 & 0.941 \\
48\_4 &      & 6.90  &11.023 & 0.716 & 1.316 & 0.942 \\
\hline
60\_1 & 0.60 & 5.541 & 8.517 & 0.722 & 1.304 & 0.942 \\
60\_3 &      & 8.40 & 10.617 & 0.723 & 1.326 & 0.959\\
\hline \hline
\end{tabular}
\label{tbl:RadCorr}
\end{table}

\subsection{Experimental Configuration}

The basic configuration of the High Resolution Spectrometer (HRS) and PbF$_2$ calorimeter were described in \cite{Alcorn:2004sb} and \cite{Fuchey:2010kna,Defurne:2015kxq}, respectively.  The PbF$_2$ calorimeter was subsequently expanded to 208 crystals. A new FPGA-based digital trigger was constructed with a two-level decision logic.
The first-level trigger logic identifies a potential electron in the HRS. It was formed by the coincidence of a scintillator (S2) signal and the CO$_2$ Cherenkov signal.  The S2 scintillator signal is the OR of the signal of any of the  12 bars making up the array.  Other ancillary first-level triggers logic (e.g. random or the coincidence of a single paddle scintillator "S0" that covers the entire HRS focal plane together with a Cherenkov signal) were also available for efficiency studies.
Each signal from the 208 PbF$_2$ channels was amplified and split into two branches, with one branch going to the Analog Ring Sampler array (ARS)
\cite{Feinstein:2003vi,Druillole:2001dm}  and the other to ADCs with 60 ns integration time.
For each channel, the ARS stores 128 analog samples at rate of 1 GHz. Upon receipt of the first-level trigger,  the ADCs digitized the integrated signal while the ARS samples were frozen.
For the kinematic settings with HRS-electron count rates below $\sim 100$ Hz, the full event was digitized and recorded for every first-level logic signal.  At higher HRS count rates, a  second level or ``trigger-validation'' process was implemented in the digital trigger.  The  208 ADC signals were digitized and summed into all possible adjacent  $2\times 2$ clusters and searched for a signal above threshold.  If no cluster above threshold was found, the event was ignored, and the ARSs were live again. If a cluster above threshold was found, then the ARS signals were digitized and the full event was recorded. The ``trigger-validation'' process required 500 ns, and this represented an irreducible deadtime for every first-level trigger event.  Upon validation, all ARS channels were digitized in parallel, at a rate of 1 $\mu$s per sample, for a total digitization time of 128 $\mu$s. Events were buffered in the DAQ to minimize any deadtime from data transfer to the main disk storage.  From there, data were periodically transferred to the central tape storage.

\begin{figure}[!ht]
  \centering\includegraphics[width=0.95\linewidth]{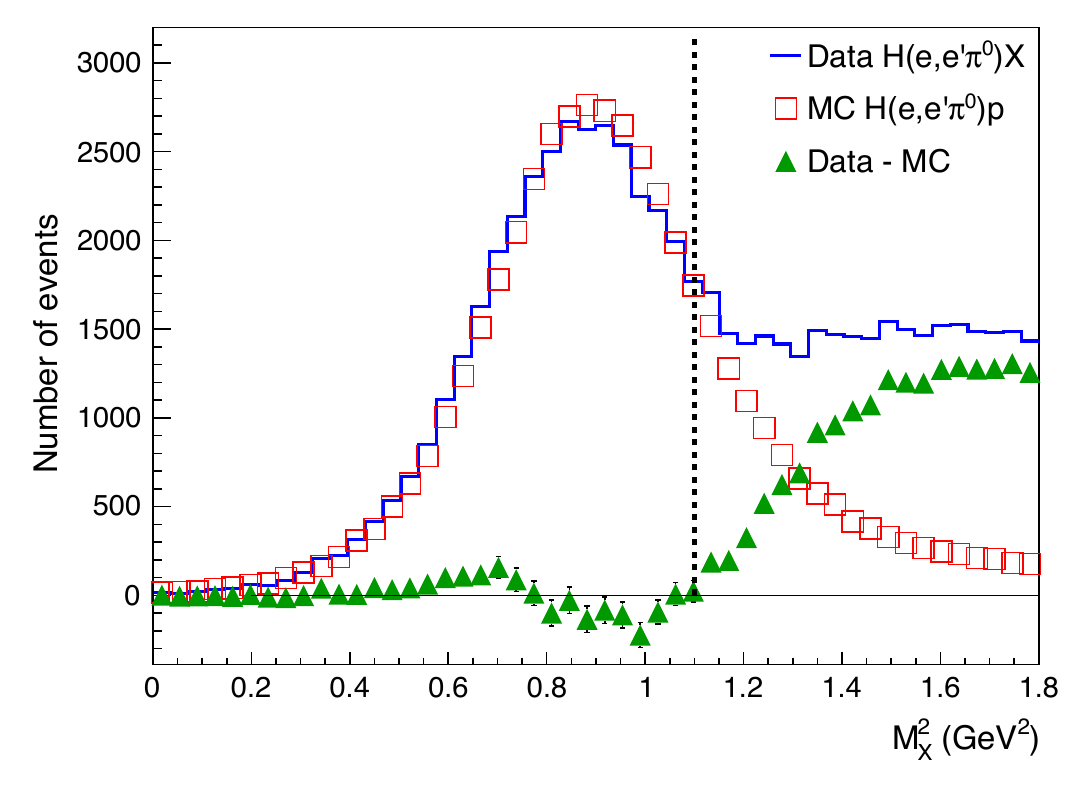}
  \caption{Missing-mass squared for $H(e,e'\pi^0)X$ events. The deviation of the experimental yield from the exclusive $\pi^0$ simulated data show the contribution of inclusive channels above the additional pion production threshold at $(M_p+m_{\pi^0})^2\approx 1.15$~GeV$^2$. The dotted line indicates the cut $M_X^2<1.1$~GeV$^2$ applied in order to remove this background. A cut on the invariant $\pi^{0}$ mass between 105 MeV and 165 MeV was also applied to ensure exclusivity.
  }
  \label{FigExclusivity}
\end{figure}

For all kinematic settings,  inclusive Deeply Inelastic Scattering (DIS) data for which the second level trigger validation was bypassed were taken simultaneously with the main DVCS/$\pi^0$ sample. The DIS triggers were prescaled to limit deadtime.   These ancillary measurements allowed to benchmark this analysis against well-known absolute cross-sections.  The DIS cross-sections extracted from these ancillary data were 4\% lower on average than the empirical fit described in Ref.~\cite{Christy:2007ve}. This systematic deviation of the DIS data from the reference is believed to arise from the time slewing of the signal from a large HRS scintillator paddle (S0), which was used in the ancillary DIS trigger but not in the main HRS-calorimeter coincidence trigger which used the S2 scintillators. Furthermore, the DIS cross-sections extracted from these ancillary data had a $\pm$4\% spread among the different kinematics settings. The estimated individual systematic uncertainties associated with luminosity, electron tracking efficiency, HRS acceptance, and acquisition deadtime added in quadrature to 3.5\%. The reported precision of the model ~\cite{Christy:2007ve} against which these measurements are compared is 2\%.  Both these sources of uncertainty explained the 4\% spread obtained in the comparison.  
The total systematic uncertainty of the $\pi^0$ electroproduction cross-section measurements includes the uncertainty on the electron detection and acceptance, the luminosity evaluation as well as the uncertainty on the photon detection and the exclusivity selection criteria. 
Figure \ref{FigExclusivity} shows how the exclusivity of the reaction  is ensured by reconstructing the missing mass  squared $M^2_X$ of the $p(e,e' \pi^0)p$ reactions.

\subsection{Extracted Structure Functions} 
\begin{figure}[!ht]
\includegraphics[bb=15 10 545 700,clip,width=\linewidth]{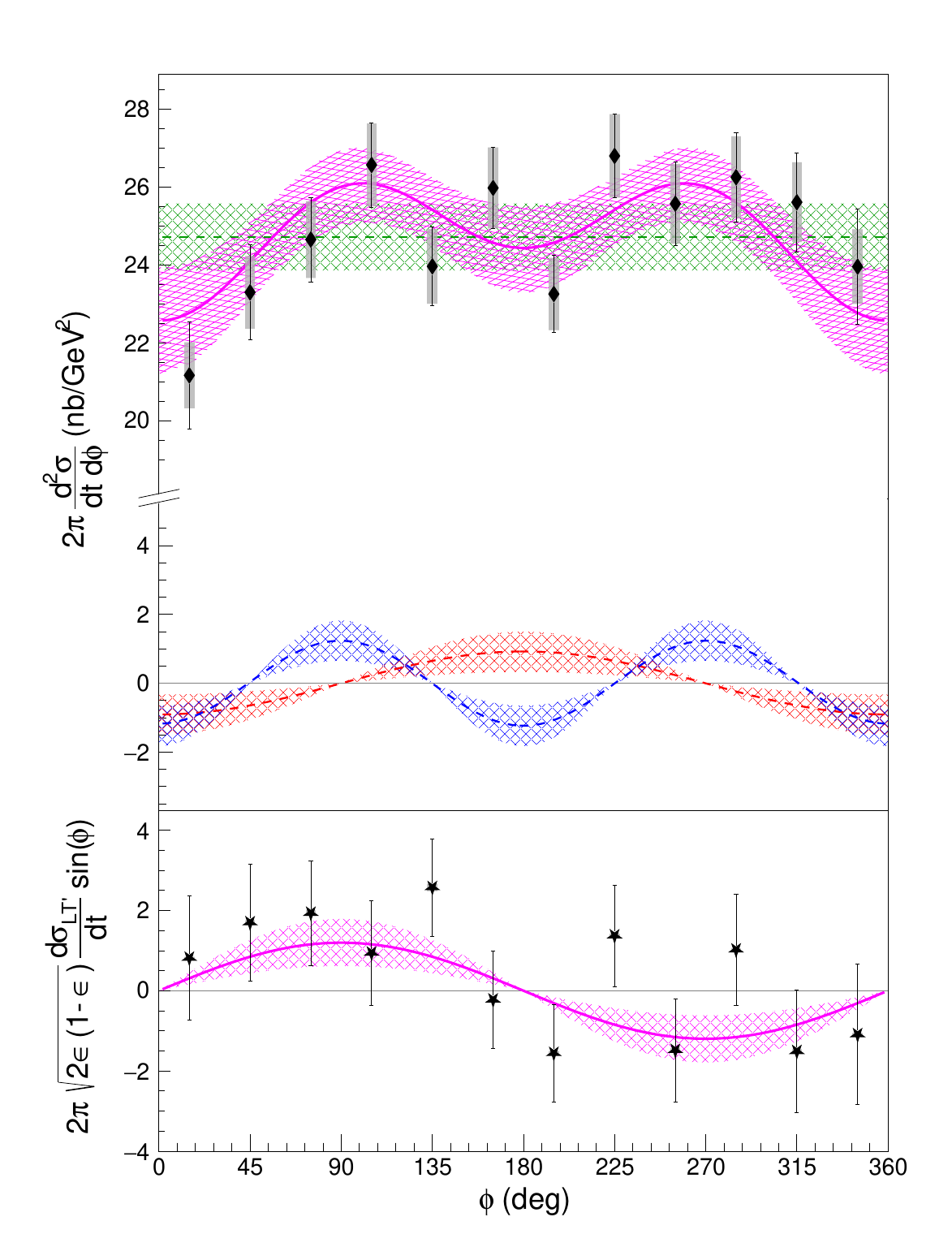}
 \caption{ 
Helicity-independent (top) and helicity-dependent (bottom) 
structure functions at $Q^2= 8.31\text{ GeV}^2$, $t'=0.15\text{ GeV}^2$, and $x_B =0.60$. The black points are the data and the magenta bands show the fits. The bars around the points indicate the statistical uncertainties, whereas the gray boxes represent the systematic uncertainties.  Contributions arising from  each of the individual structure functions of  Eq.~\ref{DVMPCrossEquation}  are also shown:  $d\sigma_U=d\sigma_T+\epsilon d\sigma_L$ in green, $d\sigma_{TT}$ in blue, $d\sigma_{LT}$ in red and $d\sigma_{LT'}$ in magenta in the bottom panel.}
\label{Figfitting} 
\end{figure}

The different structure functions appearing in the  $\pi^0$ electroproduction cross section were extracted by exploiting their specific $\phi$ dependencies and  minimizing the $\chi^2$ between the number of experimental and simulated events. Figure \ref{Figfitting} is an example of such a fit.

Table \ref{allkinparams} shows the value of the structure functions extracted in this experiment. 

\clearpage

\onecolumngrid
\begin{table*}[!hb]
\centering
\begin{tabular}{ccccccc}
\hline \hline
 & $[t'_{low},t'_{up}]$ & $\langle t' \rangle$	& $\sigma_{U}$  &   $\sigma_{LT}$   &   	$\sigma_{TT}$  &	$\sigma_{LT'}$  \\
  & ($\text{GeV}^2$) & (GeV$^2$) & (nb/GeV$^2$)  & (nb/GeV$^2$)  & (nb/GeV$^2$)  & (nb/GeV$^2$) \\
\hline

${x_{B} = 0.36}$ & $[0.00,0.07]$ & 0.03 & 195.46 $\pm$ 3.66 $\pm$ 11.93 & 7.05 $\pm$ 3.19 $\pm$ 0.25 & -4.66 $\pm$ 7.62 $\pm$ 0.16 & 14.52 $\pm$ 6.91 $\pm$ 0.51\\	

$Q^{2}=3.11$ & $[0.07 , 0.13]$  & 0.10 & 217.29 $\pm$ 4.22 $\pm$ 13.26 & -5.97 $\pm$ 3.81 $\pm$ 0.21 & -67.45 $\pm$ 9.03 $\pm$ 2.36 & 18.91 $\pm$ 8.05 $\pm$ 0.66	\\

& $[0.13 ,0.22] $& 0.18 & 195.76 $\pm$ 4.15 $\pm$ 11.95 & -11.55 $\pm$ 4.01 $\pm$ 0.40 & -67.67 $\pm$ 8.97 $\pm$ 2.37 & 27.63 $\pm$ 7.17 $\pm$ 0.97\\

 & $[0.22 , 0.38] $ & 0.29 & 183.18 $\pm$ 4.54 $\pm$ 11.18 & -28.08 $\pm$ 4.63 $\pm$ 0.98 & -87.12 $\pm$ 10.23 $\pm$ 3.05 & 9.05 $\pm$ 6.38 $\pm$ 0.32	\\

\hline
${x_{B} = 0.36}$ & $[0.00,0.06] $ & 0.03 & 115.04 $\pm$ 2.53 $\pm$ 4.64 & -2.37 $\pm$ 2.18 $\pm$ 0.08 & -16.48 $\pm$ 5.25 $\pm$ 0.58 & -5.08 $\pm$ 4.97 $\pm$ 0.18 \\
${Q^{2}=3.57}$ & $[0.06,0.13]$ &0.10 & 117.51 $\pm$ 2.77 $\pm$ 4.74 & -9.70 $\pm$ 2.54 $\pm$ 0.34 & -46.96 $\pm$ 5.85 $\pm$ 1.64 & 18.11 $\pm$ 5.31 $\pm$ 0.63 \\
& $[0.13,0.22] $ &0.17 & 119.61 $\pm$ 3.35 $\pm$ 4.82 & -5.32 $\pm$ 3.52 $\pm$ 0.19 & -35.39 $\pm$ 6.99 $\pm$ 1.24 & 14.58 $\pm$ 5.31 $\pm$ 0.51	\\
& $[0.22,0.35] $ &0.28 & 105.37 $\pm$ 4.06 $\pm$ 4.25 & -15.08 $\pm$ 4.61 $\pm$ 0.53 & -56.36 $\pm$ 8.24 $\pm$ 1.97 & 10.04 $\pm$ 4.88 $\pm$ 0.35	\\
\hline
${x_{B} = 0.36}$ & $[0.00,0.06] $ & 0.03 & 57.04 $\pm$ 1.88 $\pm$ 2.08 & -1.84 $\pm$ 1.44 $\pm$ 0.06 & -2.42 $\pm$ 3.43 $\pm$ 0.08 & 4.92 $\pm$ 3.24 $\pm$ 0.17	\\
${Q^{2}=4.44}$ & $[0.06,0.13] $ & 0.09 & 62.86 $\pm$ 2.16 $\pm$ 2.29 & -0.23 $\pm$ 1.83 $\pm$ 0.01 & -13.17 $\pm$ 4.03 $\pm$ 0.46 & 5.04 $\pm$ 3.63 $\pm$ 0.18\\
& $[0.13,0.21] $ &0.17 & 64.53 $\pm$ 2.47 $\pm$ 2.35 & 0.62 $\pm$ 2.35 $\pm$ 0.02 & -13.49 $\pm$ 4.66 $\pm$ 0.47 & 6.39 $\pm$ 3.65 $\pm$ 0.22	\\
& $[0.21,0.38] $ &0.28 & 51.63 $\pm$ 2.56 $\pm$ 1.88 & -5.66 $\pm$ 2.61 $\pm$ 0.20 & -28.80 $\pm$ 4.76 $\pm$ 1.01 & 5.79 $\pm$ 3.13 $\pm$ 0.20	\\
\hline\hline
${x_{B} = 0.48}$ & $[0.00,0.03] $ & 0.01 & 525.95 $\pm$ 14.48 $\pm$ 41.16 & 25.07 $\pm$ 16.53 $\pm$ 0.88 & -29.14 $\pm$ 43.88 $\pm$ 1.02 & 7.60 $\pm$ 30.21 $\pm$ 0.27	\\
${Q^{2}=2.67}$& $[0.03,0.06] $ & 0.04 & 520.40 $\pm$ 16.36 $\pm$ 40.73 & -38.25 $\pm$ 19.21 $\pm$ 1.34 & -7.88 $\pm$ 45.79 $\pm$ 0.28 & -5.32 $\pm$ 31.83 $\pm$ 0.19 \\
& $[0.06,0.11] $ &0.08 & 488.33 $\pm$ 17.33 $\pm$ 38.22 & -31.60 $\pm$ 21.71 $\pm$ 1.11 & -55.44 $\pm$ 47.02 $\pm$ 1.94 & 16.70 $\pm$ 28.69 $\pm$ 0.5	\\
& $[0.11,0.18] $ & 0.14 & 480.77 $\pm$ 23.45 $\pm$ 37.63 & -60.20 $\pm$ 30.66 $\pm$ 2.11 & -116.12 $\pm$ 57.67 $\pm$ 4.06 & 14.05 $\pm$ 27.05 $\pm$ 0.49	\\
\hline
${x_{B} = 0.48}$ & $[0.00,0.05] $ & 0.02 & 126.23 $\pm$ 3.84 $\pm$ 6.71 & -3.93 $\pm$ 3.36 $\pm$ 0.14 & -17.69 $\pm$ 7.80 $\pm$ 0.62 & 16.81 $\pm$ 8.43 $\pm$ 0.59	\\
${Q^{2}=4.06}$ & $[0.05,0.10] $  & 0.07 & 128.70 $\pm$ 4.65 $\pm$ 6.84 & -9.18 $\pm$ 4.45 $\pm$ 0.32 & -13.90 $\pm$ 8.66 $\pm$ 0.49 & 26.38 $\pm$ 8.65 $\pm$ 0.92 \\
& $[0.10,0.16] $ &0.12 & 115.22 $\pm$ 6.01 $\pm$ 6.12 & -16.42 $\pm$ 6.24 $\pm$ 0.57 & -23.10 $\pm$ 10.78 $\pm$ 0.81 & 30.12 $\pm$ 8.41 $\pm$ 1.05	\\
& $[0.16,0.23] $ &0.19 & 111.89 $\pm$ 8.46 $\pm$ 5.95 & -18.01 $\pm$ 8.97 $\pm$ 0.63 & -40.59 $\pm$ 13.09 $\pm$ 1.42 & 7.79 $\pm$ 8.51 $\pm$ 0.27		\\
\hline
${x_{B} = 0.48}$ & $[0.00,0.05] $ & 0.03 & 70.45 $\pm$ 2.53 $\pm$ 2.47 & 0.04 $\pm$ 2.23 $\pm$ 0.00 & -4.31 $\pm$ 5.55 $\pm$ 0.15 & 2.63 $\pm$ 4.28 $\pm$ 0.09	\\
${Q^{2}=5.16}$ & $[0.05,0.11] $ & 0.08 & 72.98 $\pm$ 2.78 $\pm$ 2.55 & 1.96 $\pm$ 2.64 $\pm$ 0.07 & -13.84 $\pm$ 5.89 $\pm$ 0.46 & 7.15 $\pm$ 4.40 $\pm$ 0.25 \\
& $[0.11,0.19] $ & 0.14 & 65.77 $\pm$ 3.17 $\pm$ 2.30 & -1.82 $\pm$ 3.51 $\pm$ 0.06 & -19.81 $\pm$ 6.74 $\pm$ 0.69 & 2.16 $\pm$ 4.12 $\pm$ 0.08 \\	
& $[0.19,0.33] $ &0.25 & 58.49 $\pm$ 3.74 $\pm$ 2.05 & -4.52 $\pm$ 4.40 $\pm$ 0.16 & -22.46 $\pm$ 7.55 $\pm$ 0.79 & 6.62 $\pm$ 3.53 $\pm$ 0.23	 \\
\hline
${x_{B} = 0.48}$ & $[0.00,0.05] $  & 0.02 & 33.48 $\pm$ 1.60 $\pm$ 1.54 & -0.79 $\pm$ 1.73 $\pm$ 0.03 & 6.43 $\pm$ 4.08 $\pm$ 0.23 & 3.23 $\pm$ 2.96 $\pm$ 0.11	\\
${Q^{2}=6.56}$ & $[0.05,0.10] $  & 0.07 & 38.21 $\pm$ 2.06 $\pm$ 1.76 & -0.39 $\pm$ 2.39 $\pm$ 0.01 & 2.15 $\pm$ 5.14 $\pm$ 0.08 & 0.98 $\pm$ 3.30 $\pm$ 0.03 \\
& $[0.10,0.15] $ & 0.12 & 31.61 $\pm$ 2.35 $\pm$ 1.46 & -4.97 $\pm$ 2.95 $\pm$ 0.17 & -4.24 $\pm$ 5.56 $\pm$ 0.15 & 10.66 $\pm$ 3.11 $\pm$ 0.37	\\
& $[0.15,0.21] $ &0.18 & 43.74 $\pm$ 4.23 $\pm$ 2.02 & 11.91 $\pm$ 5.51 $\pm$ 0.42 & -4.12 $\pm$ 8.54 $\pm$ 0.14 & 6.22 $\pm$ 3.39 $\pm$ 0.22	\\
\hline\hline
${x_{B} = 0.60}$ & $[0.00,0.12] $  &0.06 & 92.48 $\pm$ 1.42 $\pm$ 4.26 & -2.18 $\pm$ 0.70 $\pm$ 0.44 & -3.29 $\pm$ 1.67 $\pm$ 0.12 & 6.61 $\pm$ 1.68 $\pm$ 0.23 \\	
${Q^{2}=5.49}$ & $[0.12,0.26] $ &0.18 & 86.89 $\pm$ 1.39 $\pm$ 4.01 & -5.12 $\pm$ 0.75 $\pm$ 1.04 & -4.28 $\pm$ 1.66 $\pm$ 0.15 & 6.47 $\pm$ 1.60 $\pm$ 0.23	\\
& $[0.26,0.44] $ & 0.34 & 78.96 $\pm$ 1.38 $\pm$ 3.64 & -6.44 $\pm$ 0.86 $\pm$ 1.31 & -17.84 $\pm$ 1.74 $\pm$ 0.62 & 5.47 $\pm$ 1.47 $\pm$ 0.19	\\
& $[0.44,0.88] $ & 0.61 & 63.43 $\pm$ 1.37 $\pm$ 2.92 & -4.87 $\pm$ 1.03 $\pm$ 0.99 & -13.65 $\pm$ 1.82 $\pm$ 0.48 & 6.66 $\pm$ 1.10 $\pm$ 0.23	\\
\hline
${x_{B} = 0.60}$ & $[0.00,0.09] $ & 0.05 & 27.29 $\pm$ 0.90 $\pm$ 1.10 & -0.36 $\pm$ 0.48 $\pm$ 0.01 & -1.72 $\pm$ 1.25 $\pm$ 0.06 & 1.18 $\pm$ 0.90 $\pm$ 0.04 \\	
${Q^{2}=8.31}$ & $[0.09,0.21] $  & 0.15 & 24.72 $\pm$ 0.86 $\pm$ 1.00 & -0.75 $\pm$ 0.48 $\pm$ 0.03 & -2.50 $\pm$ 1.18 $\pm$ 0.09 & 1.70 $\pm$ 0.83 $\pm$ 0.06	\\
& $[0.21,0.36] $ & 0.28 & 21.43 $\pm$ 0.81 $\pm$ 0.86 & -2.43 $\pm$ 0.49 $\pm$ 0.09 & -4.71 $\pm$ 1.13 $\pm$ 0.16 & 0.21 $\pm$ 0.71 $\pm$ 0.01	\\
& $[0.36,0.69] $ &0.50 & 18.48 $\pm$ 0.79 $\pm$ 0.74 & -1.20 $\pm$ 0.58 $\pm$ 0.04 & -4.33 $\pm$ 1.18 $\pm$ 0.15 & 3.10 $\pm$ 0.56 $\pm$ 0.11	\\
\hline \hline
\end{tabular}
\caption{Numerical values of the structure functions shown in Fig. 3. The first and second uncertainty values indicate the statistical and systematic uncertainties, respectively. $\langle t' \rangle$ is weighted average of events for each $t'$ bin, with upper and lower bounds given by $[t'_{low},t'_{up}]$.}
\label{allkinparams}
\end{table*}


\end{document}